\documentclass[a4paper,fleqn]{cas-dc}

\usepackage{float}
\usepackage[numbers,sort&compress]{natbib}
\usepackage{siunitx}
\usepackage{miller}
\def\tsc#1{\csdef{#1}{\textsc{\lowercase{#1}}\xspace}}
\tsc{WGM}
\tsc{QE}
\tsc{EP}
\tsc{PMS}
\tsc{BEC}
\tsc{DE}
\DeclareSIUnit\bar{bar}

\begin{document}
\let\WriteBookmarks\relax
\def\floatpagepagefraction{1}
\def\textpagefraction{.001}

\emergencystretch 3em

\shorttitle{Delineating the interplay effects of microstructure topology and residual stresses}
\shortauthors{Ganesan \& Sandfeld}

\title[mode = title]{Delineating the interplay effects of microstructure topology and residual stresses in ultrafast laser irradiated thin films}

\author[1]{Hariprasath Ganesan}[type=editor,
                        auid=000,bioid=1,
                        orcid=0000-0002-2152-5928]
\cormark[1]

\ead{h.ganesan@fz-juelich.de}
\ead[url]{www.fzj.de}

\credit{Conceptualization of this study, Methodology, Software, Writing - Original draft preparation}

\address[1]{Institute for Advanced Simulations – Materials Data Science and Informatics (IAS‑9), Forschungszentrum Juelich, Juelich, Germany}

\author[1,2]{Stefan Sandfeld}[type=editor,
                        auid=000,bioid=2]
\credit{Funding Acquisition, Project Administration, and Review}

\address[2]{RWTH Aachen University, Faculty of Georesources and Materials Engineering,
Chair of Materials Data Science and Materials Informatics, Aachen, Germany}

\cortext[cor1]{Corresponding author}

\begin{abstract}
Advanced nanodevices require high-precision machining of thin films using ultrafast lasers. 
However, thin-film fabrications cause variations in microstructure, crystallographic orientation, and residual stresses owing to coating conditions and substrate choice. 
This work investigates the complex interplay between these factors in ultrafast laser-irradiated gold (Au) thin films using a hybrid Two Temperature Model-Molecular Dynamics simulations. 
We realized microstructure-informed atomistic models with varying grain topologies (randomized vs. equiaxed), grain sizes, and residual tensile/compressive stress configurations.
Our results reveal a clear hierarchy of influence on laser-metal interaction: 1.) Microstructure configuration 2.) Topology 3.) Grain Size 4.) Crystallographic orientations. 
In fine-grained thin films, grain boundaries act as primary melting precursors, while local crystallographic orientation determines the melting extent in coarser grains. Residual tensile stresses contribute to higher melting and greater laser-induced expansion than unstrained films. Conversely, residual compressive stresses resist deformation, as deposited thermal energy is utilized to overcome lattice compression, leading to reduced expansion. 
We found that microstructure grain topology and size exert a stronger fingerprint on film expansion than the initial defect density. 
\end{abstract}


\maketitle
\section{Introduction}

Advanced nanomanufacturing techniques involving ultrafast laser processing \cite{sugioka2017progress} of metallic thin films has a range of applications, including nanostructuring, plasmonic/optical device fabrication, surface functionalization, and energy devices.  
Laser ablation is widely used as a nanoscale processing technique for controlled material removal.
At the core of these techniques lies laser-metal interaction \cite{shugaev2016fundamentals}, which deposits high energy into the target material within ultrafast time (${10}^{-9}\thinspace\si{\second}$). 
In metals, upon excitation of conduction-band electrons, the lattice subsystem is heated via electron-phonon interactions, leading to thermal transport phenomena that influence the thin-film microstructure.

Improving the quality and precision of such nanomanufacturing processes requires a deep understanding of laser-metal interactions \cite{lin2025advancing}.
On the one hand, finding optimal laser processing parameters (e.g., laser profile, pulse duration, intensity, etc) remains challenging. 
On the other hand, accounting for variations in thin-film microstructure and thermodynamic state poses further bottlenecks.
Investigations that oversimplify any of these aspects are prone to bias and lack physical realism.

The current understanding of the complexity of ultrafast laser-metal interaction is mainly possible due to some advancements in experimental imaging and characterization techniques, including time-resolved optical methods \cite{guo2019ultrafast}, ultrafast imaging of dynamics \cite{wang2022ultrafast, martin2019ultrafast}, ultrafast diffraction and spectroscopy, and transmission electron microscopy \cite{andreev2017ultrafast}. Together, these 
techniques advanced our knowledge in aspects like electron/lattice dynamics, phase transition, plasma characterization, spallation, subsurface melt flow, temperature fields, laser-induced defects, melting, and material removal. However, a holistic understanding of these processes remains challenging, and modeling efforts \cite{miloshevsky2022ultrafast} have gained significant traction in recent times.
In particular, a hybrid approach \cite{zhigilei2009atomistic} combining the continuum two-temperature model (TTM) with molecular dynamics is particularly suitable because it can simulate ultrafast processes at atomic resolution.

For a comprehensive review of atomistic simulations of laser-driven nanomanufacturing and coupled TTM modeling works, refer \cite{liu2024review,song2023critical}.
Note that ultrafast laser-metal interaction encompasses both laser (energy) and target film (material) aspects. 
Until recently, however, several studies \cite{yin2021molecular, ivanov2023atomistic, valavanis2024laser, xie2025molecular} have focused primarily on energy-related aspects: identifying optimal laser process parameters, such as laser intensity, profile, and pulse duration; thus, the insights remain limited. 
Also, some studies have focused on femtosecond laser ablation of metals and thin films, in particular \cite{amoruso2014ultrashort, lian2024atomistic,rouleau2014nanoparticle,zhang2021mechanisms,hayder2024effective}.
Thin-film fabrication often involves physical and chemical vapor deposition techniques, resulting in variations in microstructure, residual stresses, and preferred crystallographic orientation.
Thus, a rigorous physical understanding, including these contributions, becomes essential.
Our recent work \cite{ganesan2025capturing} yielded critical insights into how thin-film microstructures and crystallographic orientations influence laser-induced deformation.
In particular, we showed that enhanced defects, such as dislocations, stacking faults, and grain boundaries, affect the ablation threshold in thin films.
Consequently, the applied laser fluence results in varying degrees of laser-induced deformation \cite{olbrich2016investigation}.

To this end, this work addresses the research question of delineating the interplay between microstructural topology and residual stresses in thin films irradiated by an ultrafast laser. 
Atomistic simulation and modeling \cite{ganesan2018parallelization, ganesan2021quantifying, ganesan2025modeling} enable efficient evaluation of these factors without incurring cross-effects.
Therefore, this work encompasses a range of atomistic models with varying microstructural topologies (i.e., grain size and grain shape) and different residual stresses reported in as-deposited thin films. Section 2 discusses the Materials \& Methods, modeling assumptions, and simulation procedure, and provides them in sufficient detail. Section 3 discusses the results from the TTM-MD simulation, focusing on the interplay of thin-film microstructure and residual stresses.

\section{Materials and Methods}\label{sec:Methods}

\textbf{TTM-MD approach and Interatomic Potential:}

Classical Molecular dynamics (MD) simulates the collective behavior of atoms representing the material, evolving in time under given initial and boundary conditions. 
MD suits the spatio-temporal context of the laser-metal interaction process, i.e., atomic length ($\si{\nano\metre}$) and ultrafast time ($\si{\femto\second}$) scale.
However, irradiation of metallic thin films with ultrafast laser pulses results in a nonequilibrium thermal state consisting of hot electrons and cold lattice subsystems. 
Here, we combined the continuum two-temperature model (TTM) and atomistic MD to systematically simulate both short-time and long-time effects encountered during laser-metal interaction.  
For specific details on the theory, implementation of TTM-MD and related parameters, we refer the readers to the previous works, including ours \cite{norman2012atomistic, yao2022exploring,ganesan2025capturing}.

Accordingly, eq.\thinspace\ref{eqn:elecTemp} indicates the continuum TTM part for describing the energy transport within the electron subsystem using the heat diffusion equation.    
\begin{equation}\label{eqn:elecTemp}
    C_{e}\rho_{e}\frac{\partial T_e}{\partial t} = \nabla(\kappa_e \nabla T_e)-\gamma_{p}(T_e - T_a)+\gamma_{s}T_a + S 
\end{equation}
where $C_{e}$ is the specific heat, $\rho_{e}$ is the electron density, $\kappa_e$ is the thermal conductivity, $\gamma_{p}$ indicates the electron-phonon coupling coefficient, $\gamma_{s}$ indicates the electron stopping coupling parameter, and $S=-I(t)\frac{exp^\frac{-x}{l}}{l}$ refers to the applied laser source \cite{thompson2022lammps}.
Herein, the electron temperature $T_e$ dependence of both specific heat and thermal conductivity is considered \cite{ganesan2025capturing}.

We discretized the atomistic model spatially with grid cells of 120 $\times$ 8 $\times$ 8 along the X, Y, and Z directions.
The atomic part describes the ion dynamics with modified Newton's equation of motion, including a coupling term $\xi = \frac{\frac{1}{n}\sum{}_{k=1}^{n} \gamma_{p}V(T_{e} - T_{a})}{\sum_{i}^{N}m_{i}v_{i}^{2}}$ for the energy exchange between the two subsystems \cite{rutherford2007effect}.
\begin{equation}
    m_{i}\frac{d^{2}\textbf{r}_{i}}{dt^{2}} = \textbf{F}_{i} + \xi m_{i}v_{i}
\end{equation}
where $m$, \textbf{r} and \textbf{F} indicates mass, position, and force of atom \textit{i}, respectively. In the coupling term $\xi$, $n$ indicates the finite difference integration step within each MD step, whereas $N$ and $V$ indicate the number of atoms and the volume of the grid cell with fitted parameters and thermo-physical quantities \cite{yao2022exploring,ganesan2025capturing}. 

Embedded Atom Method (EAM) potentials \cite{daw1984embedded} are parameterized functions that are fitted to capture the material properties of target metals and alloys.
Herein, Au has an electronic configuration with both delocalized \textit{s} electrons and localized \textit{d} electrons. 
For a faithful simulation of the laser ablation process, it is important to capture the electron temperature dependence on the physical properties of the laser ions.
To this end, we chose the Norman electron temperature-dependent potential \cite{norman2012atomistic} for the target Au thin-film system.  

\begin{figure*}
\centering
\includegraphics[scale=.65,trim={4cm 3.5cm 5cm 5cm},clip=false]{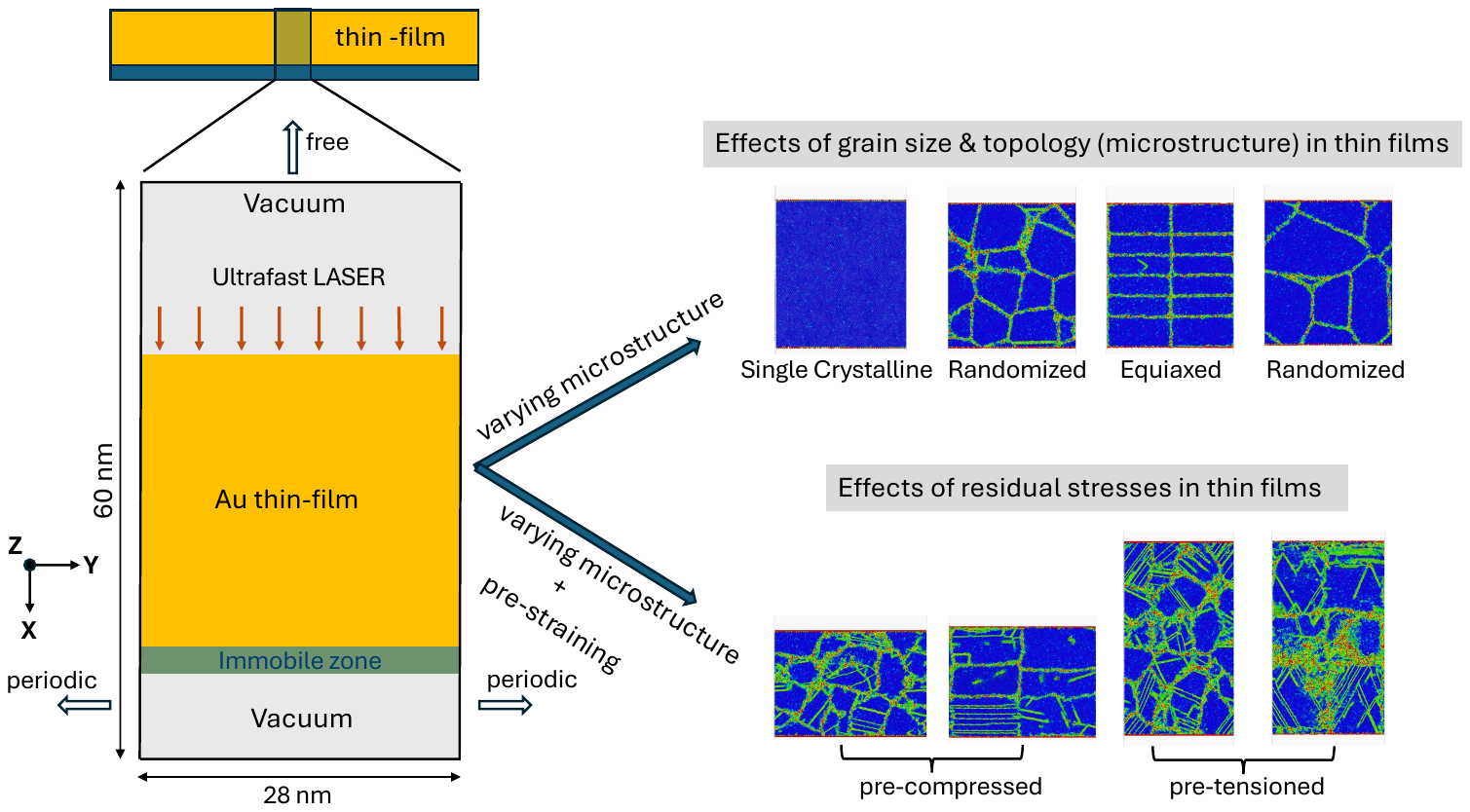}
\caption{Schematic of the atomistic model setup of Au thin-film irradiated by the ultrafast laser pulse. Consideration of two categories of thin films: 1.) Microstructure: Effects of grain size (coarser vs. finer) \& topology (Single Crystalline, equiaxed, rand), 2.) Effects of residual stresses.
\label{Fig:AtomisticModel}}
\end{figure*}  

\textbf{Atomistic modeling of thin films:}
Thin films refer to solid material layers ranging from a few nanometers to micrometers in thickness, often fabricated using physical or chemical vapor deposition techniques. 
Here, we realized atomistic Au thin-film models coated on a pseudosubstrate (i.e., vacuum), representing free-standing thin films to capture the sole contribution of material aspects under ultrafast-laser irradiation.

Figure \ref{Fig:AtomisticModel} shows the schematic of the atomistic model considered in this work for investigating ultrafast laser-metal interactions. 
Herein, the initial unstrained atomistic models have comparable dimensions of 
60$\thinspace\si{\nano\metre}$ $\times$ 28$\thinspace\si{\nano\metre}$ $\times$ 28$\thinspace\si{\nano\metre}$ (\textit{along the X, Y, and Z}) made of $\approx$ 1.6 $\times 10^{6}$ Au atoms existing in a face-centered cubic (fcc) configuration with \textit{a} = 4.08\thinspace\AA.
Unlike typically assumed rod-shaped atomistic models, we considered topologically affine Microstructure-Informed Atomistic Models (MIAMs) \cite{ganesan2021understanding,chandran2024studying,ganesan2025capturing} to incorporate microstructure variations in sufficient detail.
Owing to the choice of substrate and coating conditions, the deposited thin films are susceptible to variation in microstructure, crystallographic orientations, and residual stresses. 
Accordingly, we considered thin-film  \cite{grovenor1984development,olbrich2020hydrodynamic} with microstructure variation in 1.) \textbf{grain topology:} (Single Crystalline (SC), i.e., \hkl(-1 1 0), randomized poly-nanocrystalline (poly-NC), and equiaxed poly-nanocrystalline configurations) 
and 2.) \textbf{granularity:} (coarser (12 Grains), moderate (24 Grains), and fine-grained variants (48 Grains) of both randomized and equiaxed poly-NC  configurations. 
Here, we started with a modeled thin film rather than simulating the deposition process; thus, the roughness parameter is not significant.
All the model constructions were realized using Atomsk \cite{hirel2015atomsk} and in-house codes. To this end, the constructed free-standing atomistic Au thin-film models encompass varying microstructures, including grain topology and grain size distributions. 

\textbf{Preparation Simulation:} We performed energy minimization of these atomistic thin-film models, followed by pressure and thermal equilibration at 0\thinspace\si{\bar} and 300\thinspace\si{\kelvin}. 
A short microcanonical dynamic simulation of such equilibrated models ensured the right starting conditions.
Later, to systematically capture the contributions of residual stress, we numerically prestrained thin films to various extents under uniaxial tension and compression along the X-direction at a strain rate of $10^{9}/\si{\second}$. 
Note that such prestrained models represent thin films with varying degrees of residual stress (i.e., tension or compression), as typically reported in experiments along the lateral directions. Our simulation models take the normal direction (i.e., X) for residual stress consideration. 
We assumed periodic boundary conditions in all directions for both equilibration and prestraining simulations along with a timestep of 1\thinspace\si{\femto\second}.

\textbf{Process Simulation:}
To simulate ultrafast laser irradiation and microstructure evolution in thin films, we adopted a two-stage approach following the electron-ion equilibration time \cite{ivanov2003combined, norman2012atomistic} with further justification following the earlier works \cite{iabbaden2022molecular,ganesan2025capturing}, i.e., stage-1 with TTM-MD and stage-2 with MD, along with chosen parameters.
Accordingly, in stage-1, we simulated uniform ultrafast laser irradiation using a hybrid TTM-MD approach for 30\thinspace\si{\pico\second} with the following laser parameters (pulse duration, $\tau$ = 100\thinspace\si{\femto\second}, wavelength, $\lambda$ = 400\thinspace\AA, and laser penetration depth, \textit{l}$_{skin}$ = 20\thinspace\AA). 
Then in stage-2, we simulated microstructure/thermodynamic evolution using MD for another 70\thinspace\si{\pico\second} to capture heat transfer, melting, and phase transformation.

Herein, we applied the open-source software LAMMPS \cite{thompson2022lammps} (stable version August 2023) for all classical MD and hybrid TTM-MD simulations. 
We visualized the atomistic configurations and performed relevant characterization/analysis (i.e., Centro-symmetry parameter (CSP), phase transformation, and displacement analysis) using OVITO \cite{stukowski2009visualization}.  
 
\section{Results and Discussions}\label{sec:Results_Discussions}

This section discusses the results of Au thin films with varying microstructure, residual stresses, and their interplay during ultrafast laser metal interaction, following the two-stage simulation protocol described in section \ref{sec:Methods}.

\subsection{Effects of thin-film microstructure:}\label{sec:filmMicrostructureEffects}

To this end, we irradiated different thin film atomistic models with varying microstructural features, namely, grain topology (random vs. equiaxed) and granularity (coarser vs. finer), all evolving under a comparable applied laser fluence.
We computed the centrosymmetry parameter (CSP) for all atoms in unstrained thin-film models with varying microstructure during and after irradiation with an ultrafast single-pulse laser with an applied fluence of 1.28\thinspace$\si{\joule/\centi\meter}^2$, as we are focusing below the ablation threshold \cite{ganesan2025capturing}.

\begin{figure*}[H]
 \centering
 \includegraphics[scale=.65,trim={1cm 4.1cm 1cm 3.2cm},clip=true]{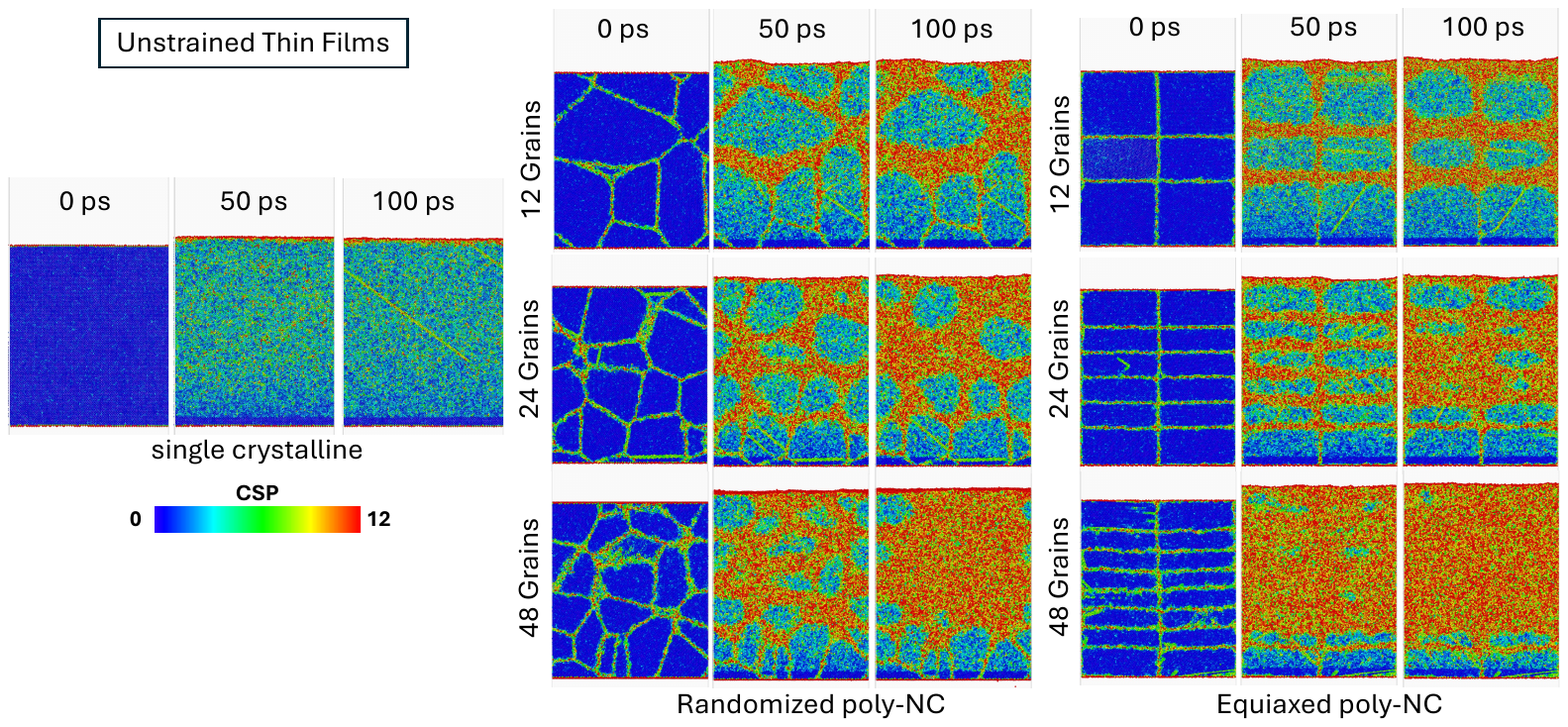}
 \caption{Comparison of Au thin-film microstructure effects: single crystalline and poly-nanocrystalline models following ultrafast laser irradiation at selected time steps. Atoms are color-coded after the centro-symmetry parameter (CSP).}\label{fig:LaserFluenceComparison}
\end{figure*}

At 0\thinspace\si{\pico\second}, the defect-free SC thin-film with \thinspace\hkl(0 0 1) orientation showed 98 \% of atoms with zero CSP values except the surface (see Fig. \ref{fig:LaserFluenceComparison}). 
Following laser irradiation, the thin-film underwent heating and expansion around 50\thinspace\si{\pico\second}, with increasing lattice distortion resulting in stacking faults around 100\thinspace\si{\pico\second}. 
Eventually, such defect-starved thin films exhibited minimal heterogeneous melting confined to the surface, as most of the laser-deposited thermal energy was utilized to overcome the configuration's cohesive energy, resulting in negligible laser-induced deformation. 

To capture thin-film grain-size sensitivity, we considered a poly-NC configuration with varying granularity of 12 (coarser), 24 (moderate), and 48 (finer).
Unlike the SC thin film, the poly-NC configuration showed relatively more grain-boundary-mediated melting.
In particular, the fine grain configurations (i.e., 48 grains) exhibit more melting regions owing to high grain boundary volume fraction. 
However, for coarser grains, the influence of local crystallographic orientation becomes dominant, leading to spatially varying melting extents.
In comparison, equiaxed fine grained thin films recorded extensive melting than randomized poly-NC counterpart, emphasizing the combined role of grain topology and grain size.

\begin{figure*}[H]
%\centering
\includegraphics[scale=.67,trim={1.8cm 4.7cm 1.5cm 3.5cm},clip=true]{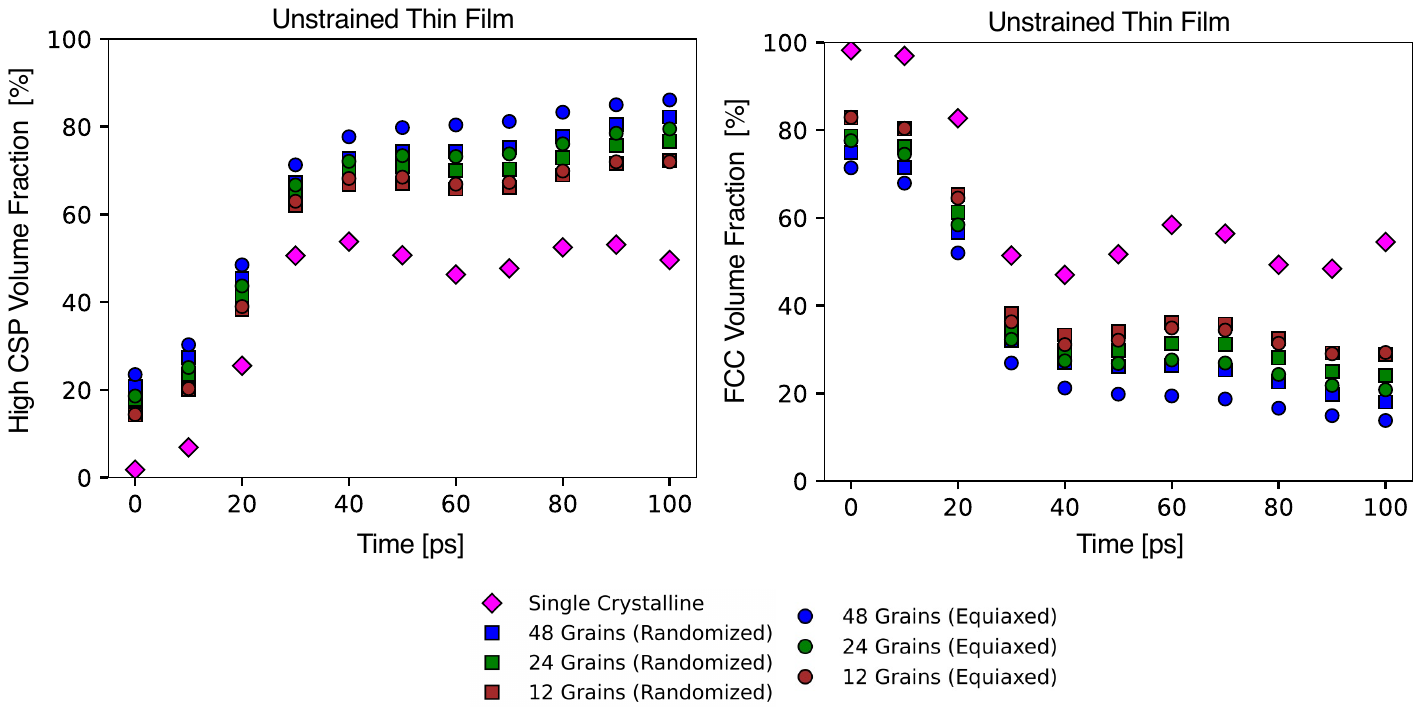}
\caption{High CSP and FCC volume fraction evolution for unstrained thin-film with varying microstructure irradiated by ultrafast laser pulse.\label{Fig:VolFractionComparison_UnstrainedThinFilm}}
\end{figure*}  

Figure \ref{Fig:VolFractionComparison_UnstrainedThinFilm} shows the temporal evolution of atoms with high CSP\footnote{for calculation, only atoms with CSP > 2.95 were considered, indicating loss of crystallinity} and FCC volume fraction for unstrained Au thin-film with varying microstructure. Herein, the SC configuration started from nearly 0\% and rose to 50\% and fluctuated slightly owing to tension and compression stages, whereas all poly-NC  configurations showed a nonlinear increase in high CSP volume fraction, among which equiaxed configurations with (finer grains) recorded relatively higher CSP volume fraction throughout in comparison to the randomized configuration. 
The FCC volume fraction decreased continuously, confirming loss of crystallinity due to melting. 
For both high CSP and FCC volume fraction, single-crystalline thin films showed considerable fluctuations due to the laser-induced pressure wave, resulting in compression and subsequent tension. 
In contrast, both randomized and equiaxed poly-NC configurations with different granularity showed nearly monotonous increase (high CSP) and decrease (FCC) trend due to faster melting. However, coarser 12 grains poly-NC thin films showed almost negligible fluctuations. These results emphasize the critical role of initial grain boundary volume fraction vs. crystalline region in accommodating laser-induced stresses and the corresponding phase volume fraction change. Thus, for consistent ablation behavior, a fabricated thin film with fine-grained microstructure is deemed more suitable, owing to its high grain-boundary volume fraction.   

\begin{figure*}[H]
 \centering
 \includegraphics[scale=.65,trim={2cm 4.9cm 2cm 2.8cm},clip=true]{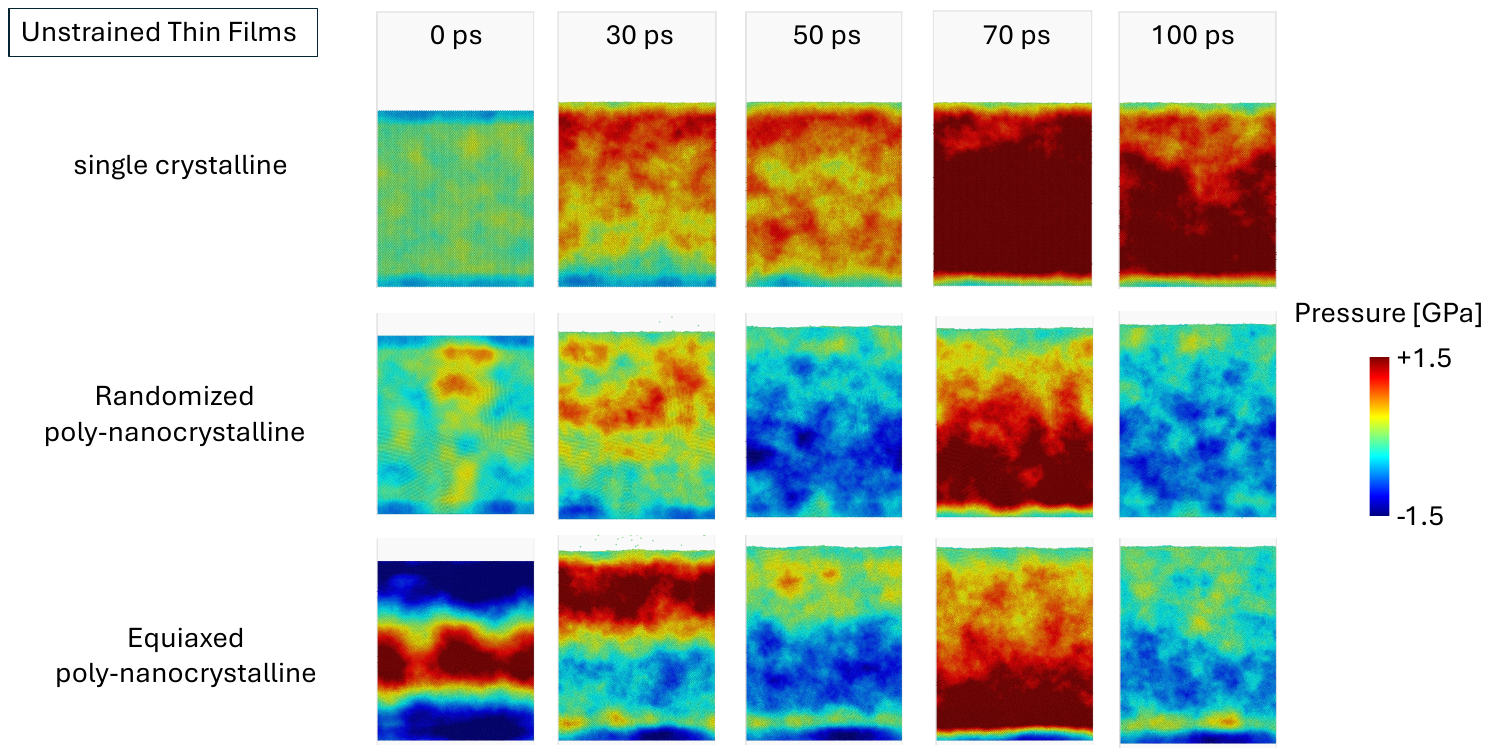}
 \caption{Spatio-temporal pressure plot comparison for unstrained thin films. For the representation of poly-nanocrystalline  configurations, we considered only thin films with 48 grains (i.e., finer).}\label{fig:PressureDistro_Unstrained_ThinFilm}
\end{figure*}

Figure \ref{fig:PressureDistro_Unstrained_ThinFilm} shows the spatio-temporal distribution of atomic pressure for unstrained thin films measured following ultrafast laser irradiation. 
Herein, SC thin-film showed an extensive pressure-free region measuring around 0\thinspace\si{\giga\pascal}, whereas the thin-film with randomized poly-NC  microstructure showed some islands of mild compression (red) and tension (blue) regions. 
For brevity, only fine-grained poly-NC configurations as representations are shown here. 
Comparing the initial distribution (at 0\thinspace\si{\pico\second}) captures the spatial pressure-gradient pattern arising from the sensitivity of underlying microstructural changes prior to laser irradiation.
Consequently, equiaxed poly-NC  thin films result in regions of extreme compression and tension. 
Our results revealed that variation in thin-film microstructural features (i.e., grain topology and grain size) strongly influenced the initial pressure distribution, thereby delineating the microstructural fingerprint's contribution to subsequent pressure evolution following ultrafast laser-metal interaction. 
In general, for SC thin films, the lack of defects and grain boundaries leads to slower compression and subsequent heating. In contrast, poly-NC  thin films exhibit faster compression and tension cycles, characterized by large variations in field pressure magnitude (between 30 and 100 \thinspace\si{\pico\second}) and by global pressure plots.    

\begin{figure*}
\centering
\includegraphics[scale=.65,trim={1.3cm 5.7cm 1cm 4cm},clip=true]{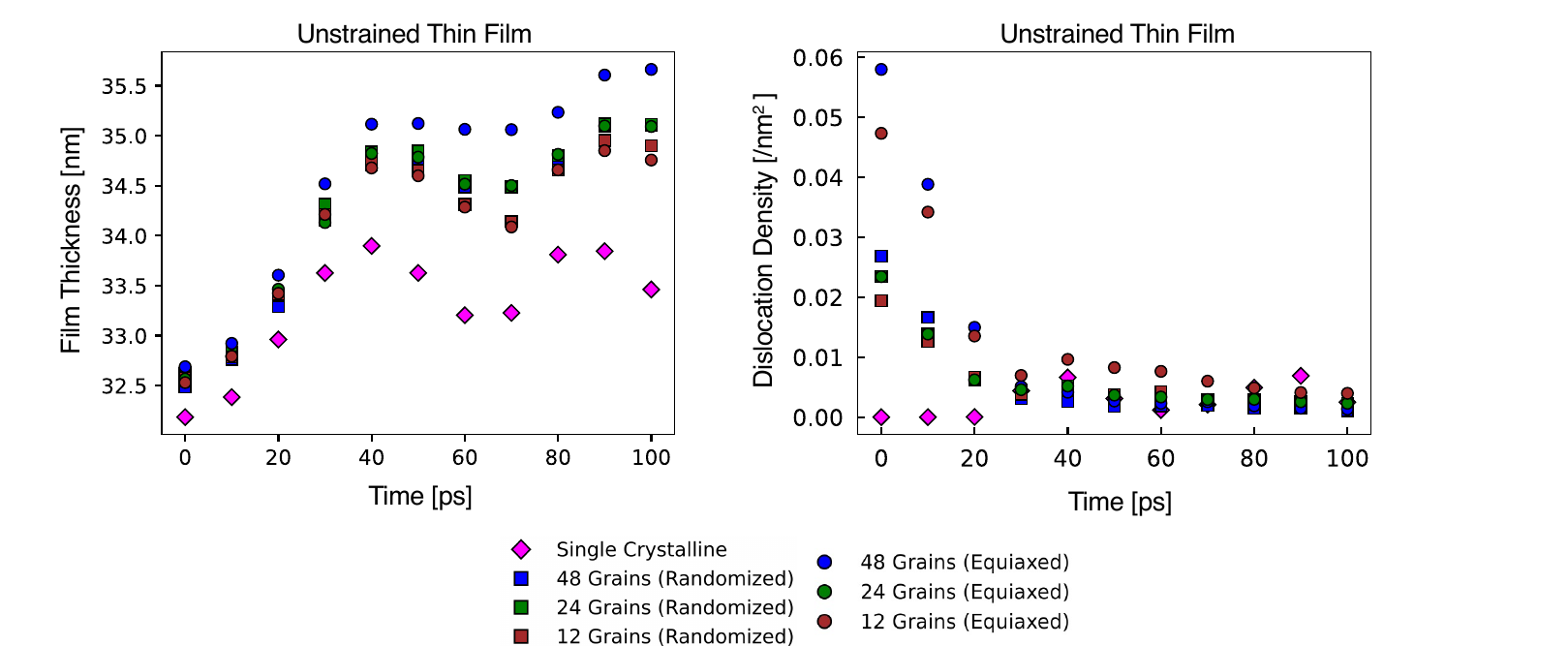}
\caption{Unstrained Film thickness and dislocation density evolution under the fluence of 1.281 \thinspace$\si{\joule/\centi\meter}^2$.\label{Fig:ThicknessDefectEvolution}}
\end{figure*}

Furthermore, we quantified geometric variation by characterizing the evolution of film thickness (see Fig.\thinspace\ref{Fig:ThicknessDefectEvolution}) for different thin-film models with varying microstructures irradiated by an ultrafast laser pulse. For SC models, the thin-film thickness increased smoothly up to 40\thinspace\si{\pico\second} and then fluctuated strongly due to compression and tension cycles, reaching a maximum thickness of 1.9 \thinspace\si{\nano\meter}. On the other hand, poly-NC  thin films showed relatively higher expansion with mild fluctuation ($\approx$ 0.5\thinspace\si{\nano\meter}).
Altogether, fine-grained (i.e., 48 grains) equiaxed poly-NC  thin films showed greater film thickness expansion ($\approx$ 3.1\thinspace\si{\nano\meter}), suggesting that the combination of grain topology and size/granularity plays a decisive role.
Additionally, in only extreme cases did preexisting defects make a weak contribution to the thin-film expansion following laser irradiation. For instance, the SC thin film, which initially had zero dislocation density, resulted in the lowest thin-film expansion compared to the equiaxed thin-film poly-NC , which started with the highest dislocation density (0.058\thinspace/$\si{\nano\meter}^{2}$). 
Most importantly, the distribution of initial dislocation density for other poly-NC  thin films did not show a strong spread ($\approx$ 0.04\thinspace/$\si{\nano\meter}^{2}$) across final film thicknesses, indicating that microstructure topology and size have a stronger fingerprint than initial defect density.

Thus, from our extensive simulations, we propose the following hierarchy of influence for thin-film microstructure: 
\begin{enumerate}
    \item Microstructure configuration (SC vs. Poly-NC)
    \item Microstructure topology (Random/Equiaxed)
    \item Granularity (coarser vs. finer)
    \item Crystallographic orientations
\end{enumerate}
It is worth emphasizing that in our earlier work \cite{ganesan2025capturing}, we clearly demonstrated the effects of grain-boundary-mediated melting (exhibiting a significant heterogeneous melting profile) in Au thin films. 

\subsection{Effects of Residual stresses in thin films:}\label{sec:EffectsOfResidualStresses}
 
Experimental investigation \cite{huff2022residual} characterized and validated the existence of residual stresses originating from the tensile
or a compressive state of thin films.
To delineate such residual stress contributions in thin films, we mechanically prestrained thin-film models and compared them with the unstrained (as-deposited) configuration.

\begin{figure*}[H]
 \centering
 \includegraphics[scale=.65,trim={0.0cm 3.6cm 0.1cm 2.0cm},clip=true]{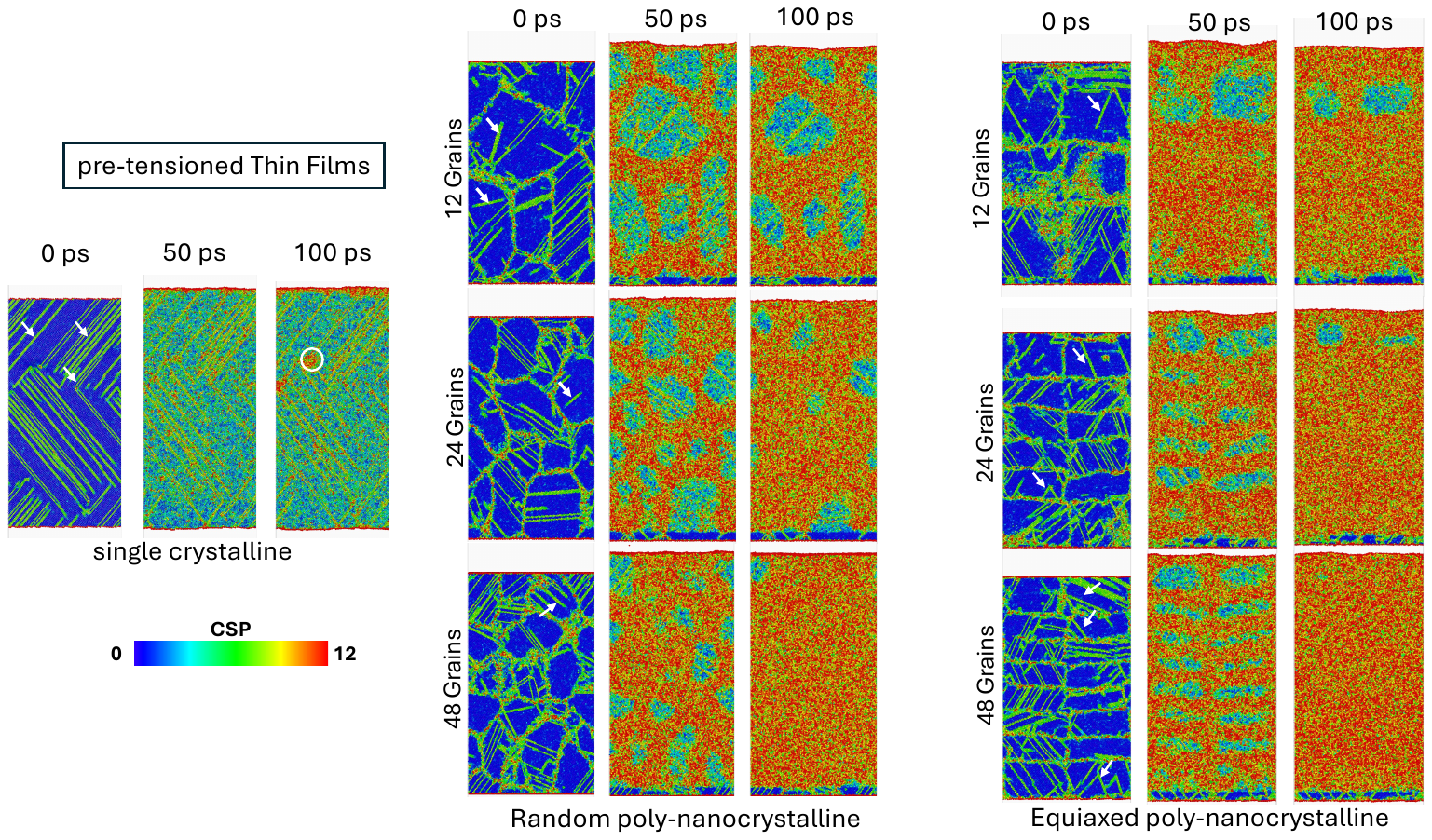}
 \includegraphics[scale=.65,trim={0cm 6.1cm 0.0cm 4.8cm},clip=true]{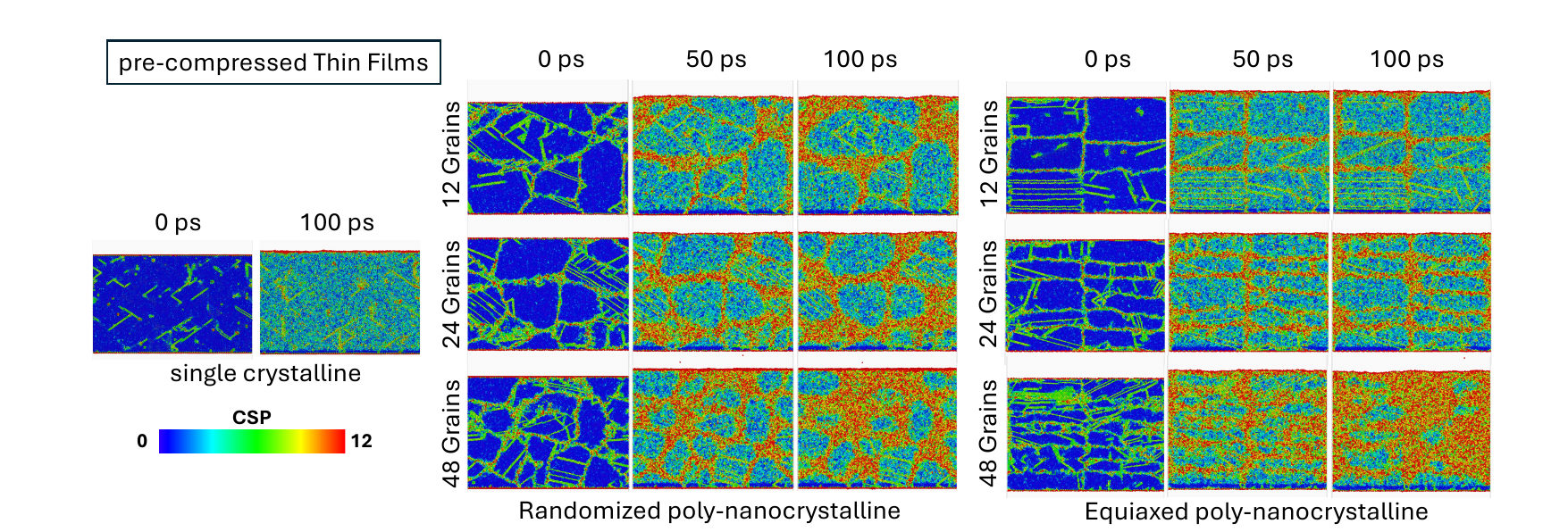}
 \caption{Comparison of prestrained (Tension vs. Compression) single crystalline and poly-nanocrystalline microstructure models irradiated with applied laser fluence. Atoms are color-coded after the centro-symmetry parameter (CSP). The white arrows indicate stacking faults, and the white circle denotes several isolated melting regions.}\label{fig:prestraining_MicroStrucComparison}
\end{figure*}

For an SC thin-film, the tensioning procedure led to the nucleation of stacking faults 
(see white arrow markers in Fig.\thinspace\ref{fig:prestraining_MicroStrucComparison}), characterized by several activated parallel slip planes. 
Following ultrafast laser irradiation, around 50\thinspace\si{\pico\second}, we observed considerable thin-film expansion, with inhomogeneous melting confined to the surface. 
Locking of cross-gliding planes results in high local lattice distortions, which facilitate melting.
Consequently, at 100\thinspace\si{\pico\second}, some melted regions (see white circle in Fig.\thinspace\ref{fig:prestraining_MicroStrucComparison}) in isolation existed deeper inside the thin-film away from the surface.   
In contrast, during pre-tension, both randomized \& equiaxed poly-NC  thin films resulted in relatively fewer stacking faults (see Fig.\thinspace\ref{fig:prestraining_MicroStrucComparison} at 0\thinspace\si{\pico\second}). 
To a large extent, both grain size and local crystallographic orientation determine the extent of stacking faults, as grain boundaries can hinder their propagation.   
Unlike unstrained poly-NC  thin films (see Fig. \ref{fig:LaserFluenceComparison}), around 50\thinspace\si{\pico\second}, we observed a slightly higher film expansion even for coarser grains characterized by grain boundary mediated melting.
Evidently, around 100\thinspace\si{\pico\second}, equiaxed poly-NC  thin-film showed a greater extent of melting than a randomized microstructure.
It is worth noting that prestraining resulted in a smaller average grain size due to lateral compression, besides nucleating new dislocations from the grain boundaries, thus, high initial defect density.
In contrast to unstrained cases, residual tensile stresses in
all thin films with varying microstructures showed higher melting and laser-induced deformation. 
This observation emphasizes the complex interplay of grain topology, grain size, residual stresses, and initial defect concentration.

\begin{figure*}[H]
%\centering
\includegraphics[scale=.67,trim={1.8cm 6.5cm 1.5cm 5cm},clip=true]{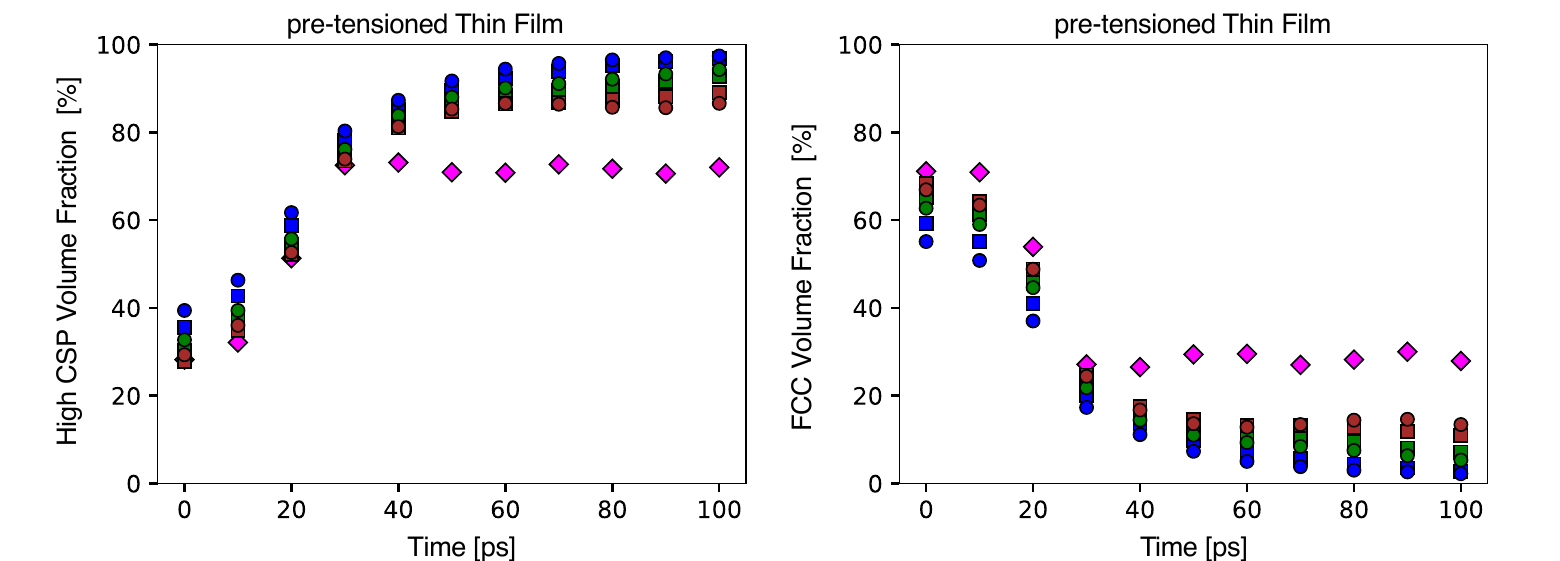}
\includegraphics[scale=.67,trim={1.8cm 4.05cm 1.2cm 4.7cm},clip=true]{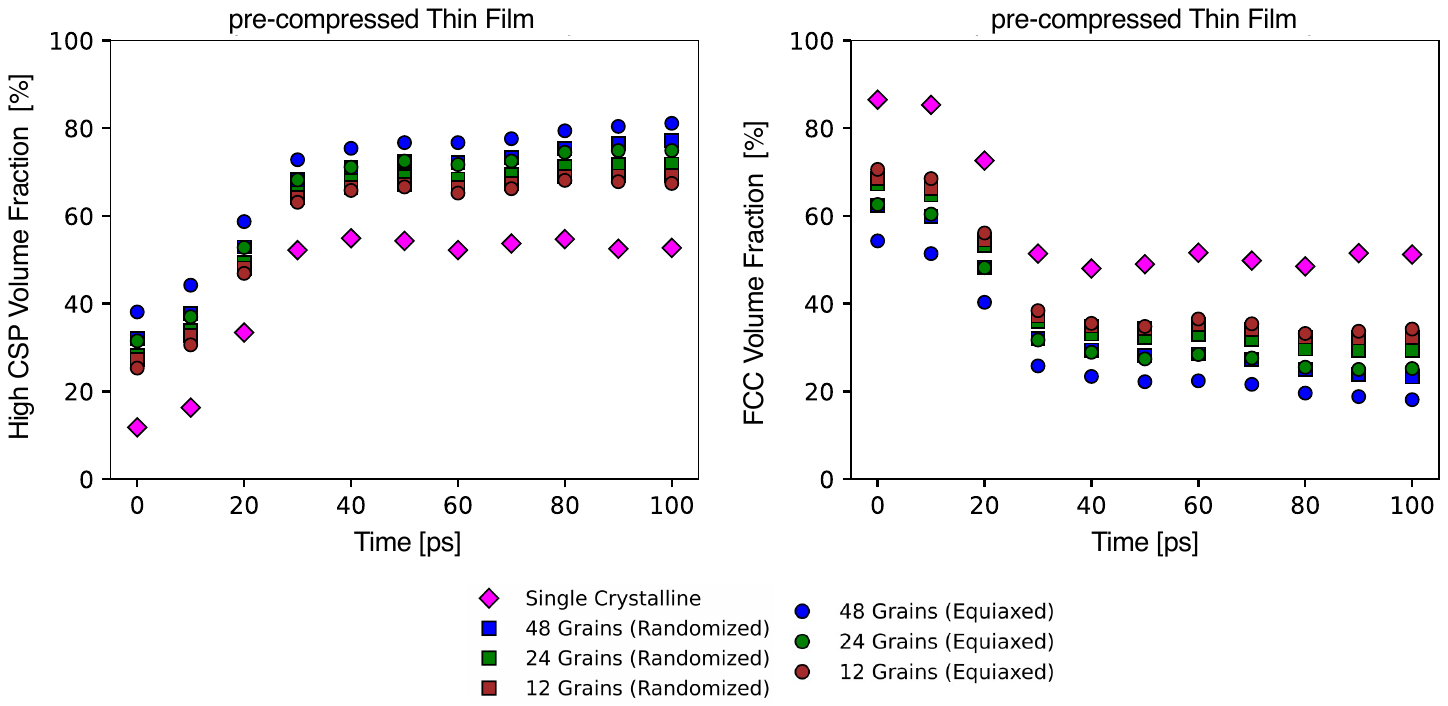}
\caption{Comparison of CSP and FCC volume fraction for pre-tensioned and pre-compressed thin films with varying microstructure (i.e., single crystalline, randomized, and equiaxed poly-nanocrystalline).\label{Fig:PreStrainingComparison}}
\end{figure*}  

In unstrained thin films, the initial high CSP volume fraction stems solely from microstructural features, whereas in tensioned thin films, deformation-induced defects also contribute to high CSP. 
Eventually, the SC thin films start from slightly high CSP values (27\%) and remain around 70\thinspace\% after 40\thinspace\si{\pico\second}. 
In contrast, finer equiaxed poly-NC  thin films showed a continuously increasing high CSP volume fraction reaching a maximum of slightly under 100\%. 
Altogether, the poly-NC  thin films showed a tight spread, indicating that the pretensioning procedure contributed significantly to laser-induced lattice distortion. 
In general, the FCC volume fraction showed a monotonically decreasing trend (\textit{consistent inverse trend when compared with high CSP}) for all tensioned thin films. In particular, the fine-grained equiaxed microstructure resulted in nearly complete melting, whereas the SC thin film fluctuated around 30\% after the initial drop.

\begin{figure*}[H]
 \centering
 \includegraphics[scale=.65,trim={2cm 4.1cm 2cm 2.8cm},clip=true]{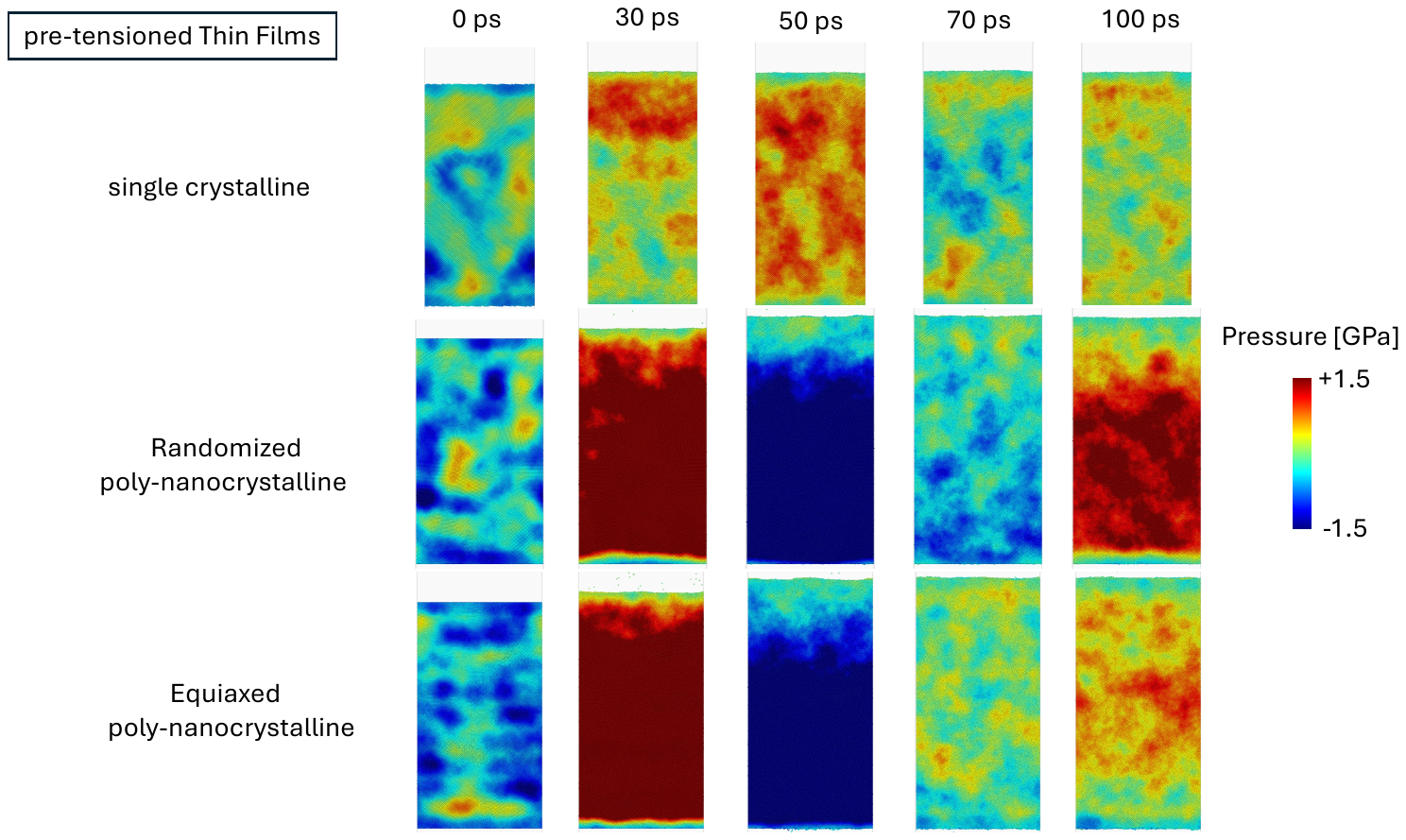}
 \includegraphics[scale=.65,trim={1.5cm 5.9cm 1.3cm 3cm},clip=true]{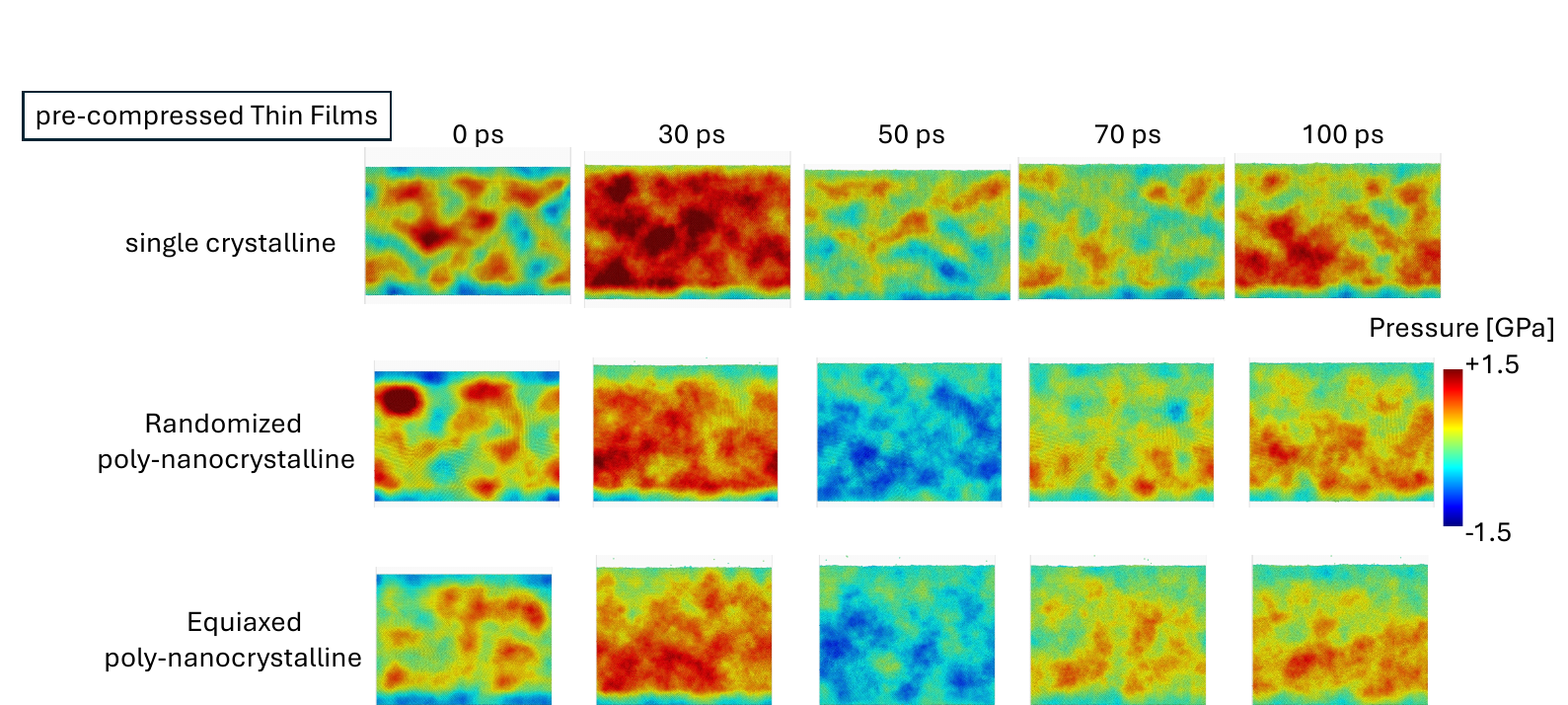}
 \caption{Spatio-temporal pressure field plot comparison for pre-tensioned and pre-compressed thin films.}\label{fig:PressureDistroComparison_prestrainedFilms}
\end{figure*}  

Pre-compressed thin-film with SC microstructure yielded moderate defect densities, with several short glided slip planes that led to locking due to cross gliding. 
Such locking junctions served as hotspots, leading to lattice distortions characterized by high CSP values.
Following ultrafast laser irradiation, these thin films exhibited negligible thin-film expansion and heterogeneous surface melting.
However, the randomized poly-NC thin-film exhibited a slightly modified grain topology, with some grains exhibiting relatively more stacking faults due to favorable crystallographic orientations (see coarser equiaxed Poly-NC thin film in Fig.\thinspace\ref{fig:prestraining_MicroStrucComparison}). 
Across different pre-compressed fine-grained polycrystalline thin films (both randomized and equiaxed), grain boundary-mediated melting was more pronounced.
In our previous investigation \cite{ganesan2025capturing} of ultrafast laser irradiation on unstrained thin films with varying microstructures, we observed an initial thin film compression to resist laser-induced lattice straining, followed by tension as a precursor to melting. 
Consequently, pre-compressed thin films tend to further reduce laser-induced thin-film expansion, as the laser-deposited thermal energy gets utilized to overcome lattice compression. 
Our results reveal that residual compressive stresses in thin films strongly reduced the extent of laser-induced deformation relative to thin-film microstructural features, namely, grain size and topology. 

The pre-compressed poly-NC configurations showed a comparable initial high CSP volume fraction (0\thinspace\si{\pico\second}) that increased gradually following laser irradiation, reaching a maximum ranging from 65-80\thinspace\%, whereas the SC film saturated around 50\thinspace\% after 30\thinspace\si{\pico\second}.
Despite starting from high FCC volume fractions (at 0\thinspace\si{\pico\second}), the poly-NC thin films gradually declined, reaching a minimum of 23\thinspace\%.
Altogether, these observations reveal that residual stresses in pre-compressed thin films exhibit strong resistance to laser-induced deformation; thus, optimization of nanostructure ablation process parameters with ultrafast lasers should account for any residual stresses in thin films. 

In Figure.\thinspace\ref{fig:PressureDistroComparison_prestrainedFilms}, after pre-tension, both SC and poly-NC  thin films exhibited a distribution of strong and moderate tension (blue) regions, along with some islands of compression (red) regions. 
Following ultrafast laser irradiation, both poly-NC  thin films underwent compression and tension cycles for up to 50\thinspace\si{\pico\second}, after which the pressure intensities decreased considerably.
Dedicated experimental investigations \cite{chen2011time, wu2022ultrafast,liubchenko2025laser} on laser irradiation of metallic thin films agree with our observations of laser-induced compressive stresses due to ultrafast heating, followed by structure change accompanied by pressure relaxation resulting in breathing of thin films. 
Specifically, our previous work \cite{ganesan2025capturing} on Au thin films revealed that defect nucleation mechanisms like cavitation, dislocations and planar defects act as pressure relieving mechanisms based on microstructure-dependent pressure threshold resulting in pressure drop. 
At 100\thinspace\si{\pico\second}, the random poly-NC  thin film exhibited slightly higher compressive values than the equiaxed counterpart due to incomplete melting.    
Following precompression, different thin films with varying microstructures (at 0\thinspace\si{\pico\second}) showed a similar distribution of high- and moderately compressed regions. 
Interestingly, SC thin films retained predominantly high compression regions following laser irradiation.
On the other hand, random/equiaxed poly-NC  thin films showed a weakly fluctuating tension and compression regions until (50\thinspace\si{\pico\second}), followed by a moderate distribution of compressive stresses around 100\thinspace\si{\pico\second} in thin films. 
Unlike pre-tensioned thin films, no strong pressure intensities were observed herein.
These observations emphasize how thin-film residual stresses influence laser-induced pressure and targeted machinability. 

\begin{figure*}
\centering
\includegraphics[scale=.65,trim={1.3cm 6.7cm 1cm 5.1cm},clip=true]{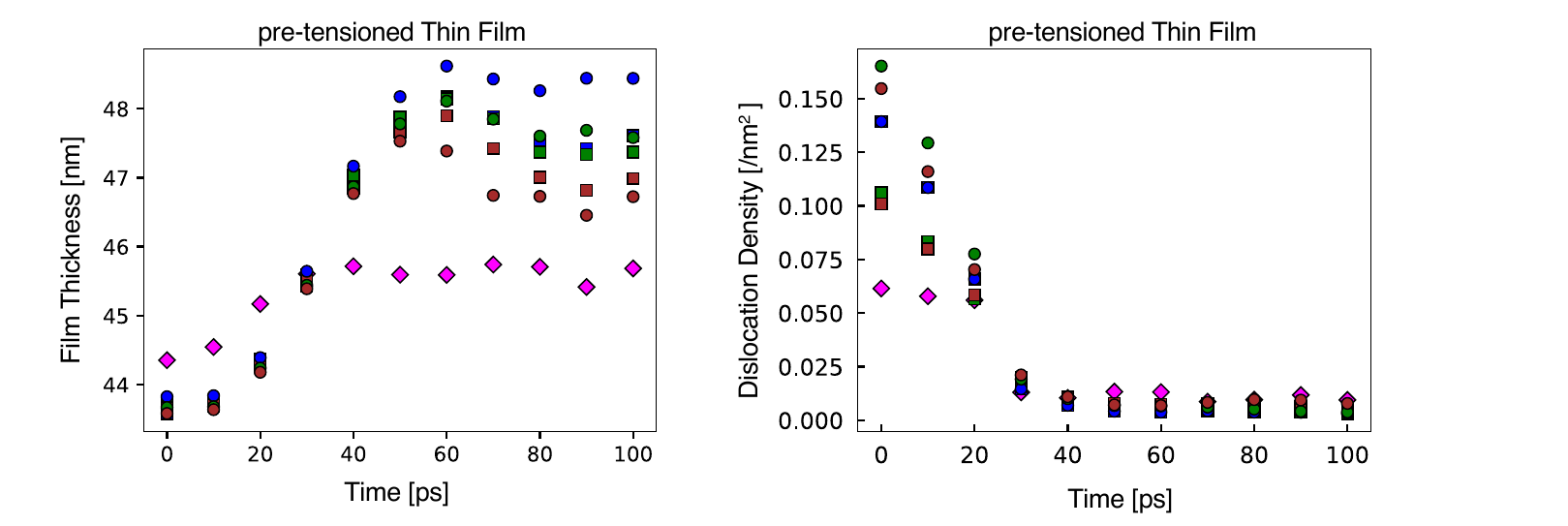}
\includegraphics[scale=.65,trim={1.35cm 4.7cm 1cm 5cm},clip=true]{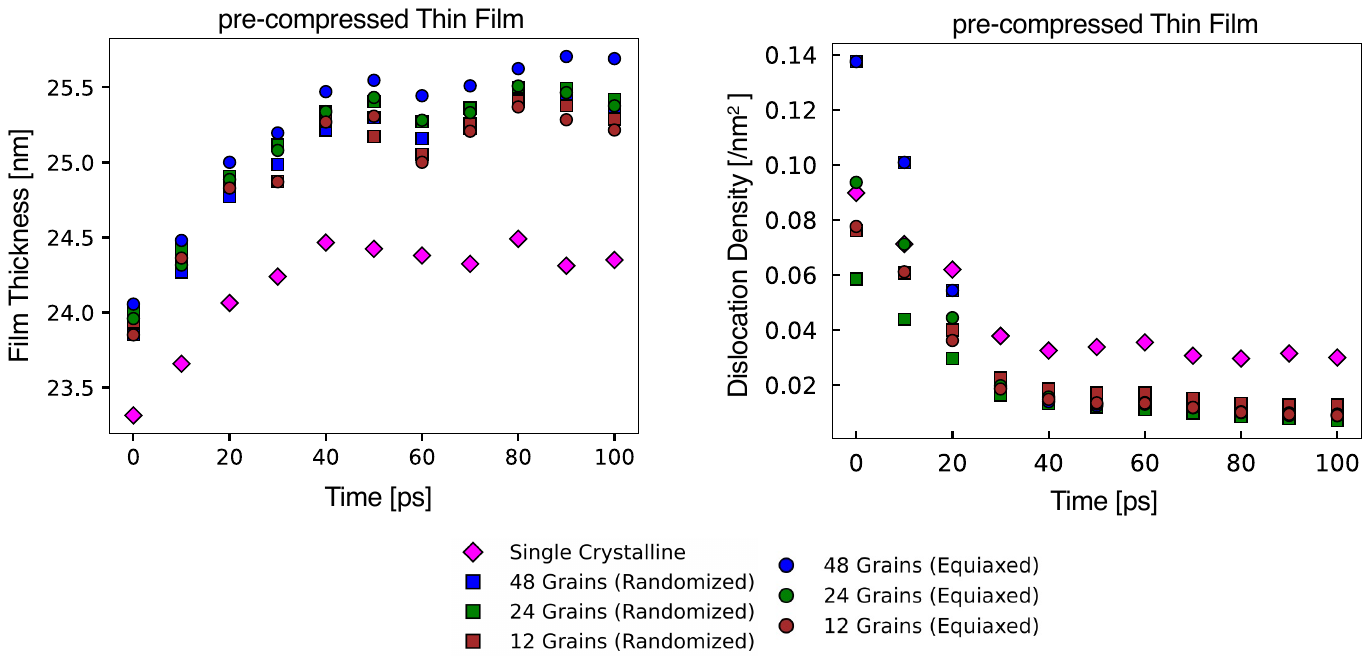}
\caption{Prestrained thin-film samples - Film thickness and dislocation density evolution under the fluence of 1.281 \thinspace$\si{\joule/\centi\meter}^2$.\label{Fig:Prestrained_ThicknessDensity}}
\end{figure*}

For an unstrained SC thin film, the peak film thickness expansion was 2.0\thinspace\si{\nano\meter}. In contrast, the pre-tensioned and pre-compressed SC thin films showed considerably higher and lower peak expansions of roughly 4.5\thinspace\si{\nano\meter} and 1.6\thinspace\si{\nano\meter} (see Fig. \ref{Fig:Prestrained_ThicknessDensity}), respectively.
Note that for the pre-tensioned case, both poly-NC  thin films started at slightly lower values than in the SC case (at 0\thinspace\si{\pico\second}) and then reached a peak around 60\thinspace\si{\pico\second}, after which they declined moderately.
Among the poly-NC  thin films, the fine-grained equiaxed case recorded a peak film expansion of nearly 5\thinspace\si{\nano\meter} in contrast to the fine-grained randomized case showing 3.4\thinspace\si{\nano\meter}.
Following precompression, both poly-NC  thin films showed a gradual increase in thickness, with mild fluctuations, eventually reaching an average peak of 1.5\thinspace\si{\nano\meter}. The distribution spreads over time after 40\thinspace\si{\pico\second}.
Comparing the two prestrained samples, the finer equiaxed polycrystalline thin films (i.e., 48 grains) exhibited the highest film thickness, underscoring the role of grain topology and grain-size interplay in thickness evolution.
Note that the prestraining procedure resulted in different initial thicknesses ($\pm 10\thinspace\si{\nano\meter}$ of unstrained film thickness as reference) of thin films that accommodate residual stresses, irradiated by an ultrafast laser pulse at a comparable applied laser fluence.
\cite{zhou2022experimental} performed experimental and numerical TTM-MD simulations of gold films with different thicknesses and reported a strong correlation between the separation threshold and increasing film thickness. However, these findings did not account for residual stresses in thin films.
Often, ultrafast laser irradiation of metallic thin films results in a spatial distribution of varying absorbed laser energy intensities (i.e., low to high).
Thus, our investigations reveal that the interplay between microstructure and residual stresses warrants consideration in the development of physically informed threshold parameters at lower applied fluence, whereas laser-induced pressure effects dominate at higher applied fluence.

Furthermore, we computed the evolution of the dislocation density ($\rho_{Dis}$) in different thin-film models irradiated by the applied laser fluence studied herein.
In prestrained thin films, defects originate from both original misfit dislocations at grain boundaries and reactions induced by the applied strain.
For the pre-tensioned case, the equiaxed poly-NC  thin films with 48 grains showed the highest initial dislocation density ($\approx 0.160 \thinspace/ \si{\nano\meter}^{2}$), followed by 24 grains. 
Relatively, coarser equiaxed thin films exhibited a higher initial dislocation density, owing to consistent, uniform activation volumes available across all grains. 
However, randomized grain topology showed a slightly different trend in starting dislocation densities, underscoring the role of local crystallographic orientation in determining activated slip planes and available activation volumes. Although the SC thin-film underwent comparable straining, the starting dislocation density was the lowest. In all cases studied herein, the dislocation densities decline sharply due to the loss of crystallinity caused by melting.  
In contrast, for the precompression case, the finer randomized thin-film with 48 grains showed a higher dislocation density of $\approx 0.140 \thinspace/ \si{\nano\meter}^{2}$, followed by SC with $\approx 0.09 \thinspace/ \si{\nano\meter}^{2}$ and finally randomized thin-film with 24 grains showing $\approx 0.06 \thinspace/ \si{\nano\meter}^{2}$. 
Note that some crystalline regions exhibit more stacking faults and dislocations under different prestraining conditions, primarily due to tension-compression asymmetry in the material's deformation mechanisms, arising from differences in slip systems and dislocation behavior. 
Furthermore, key factors, such as activated deformation mechanisms, critical resolved shear stress, and Schmid factors, are sensitive to loading direction.

\section{Conclusion}

This work advances the understanding of optimal laser process parameter selection by systematically delineating the roles of thin-film microstructure and residual stresses under ultrafast laser irradiation.
Microstructural features in thin films govern ultrafast laser-metal interaction in a specific order of importance: Microstructure configuration (Single Crystalline vs. Polycrystalline) → Topology (Random/Equiaxed) → Granularity → Crystallographic orientation.
Residual tensile stresses contributed to higher melting and greater laser-induced expansion (i.e., 5\thinspace\si{\nano\meter} for thin film with equiaxed 48 grains) than unstrained films (i.e., 2\thinspace\si{\nano\meter}).
On the other hand, residual compressive stresses resisted laser-induced deformation, as laser-deposited thermal energy was utilized to overcome lattice compression, leading to reduced lattice straining and lower film expansion (i.e., $\approx$1.5\thinspace\si{\nano\meter}).
The grain topology and granularity are more influential than the initial defect density in determining thin film expansion.
For fine-grained thin films, grain boundaries (GBs) act as the primary sites for melting initiation, whereas, in coarser grains, the local crystallographic orientation determines the spatial extent of that melting.
At lower applied fluences, the interplay between microstructure and residual stress governs the deformation of thin films, whereas laser-induced pressure effects become dominant at higher fluences.
Our results demonstrate that neglecting either microstructural topology or residual stress state in process models increases uncertainty in predicting ablation thresholds that affect machinability precision.
These findings emphasize the need for more rigorous physical models that incorporate microstructure, residual stresses, and deposited laser energy profiles to determine optimal processing conditions, i.e., to consider both energy and material aspects of ultrafast laser-metal interactions.
To this end, critical information, such as the residual stress state, along with the microstructural characterization of the fabricated thin films, aids the efficient tuning and optimization of ultrafast laser processing parameters in advanced nanomanufacturing scenarios, where thin-film microstructure and residual stress state are susceptible to change owing to the underlying substrate and processing history.   
Future studies investigating substrate choice beyond free-standing thin films, multi-pulse regimes, and machine-learning-driven microstructure optimization would further generalize these findings toward industrially relevant nanomanufacturing conditions.

\section{Declaration of Competing Interest}
The authors declare that they have no known competing financial interests or personal relationships that could have appeared to influence the work reported in this work.

\section{Acknowledgments}

The authors gratefully acknowledge funding from the German Research Foundation (DFG) - Grant No. 469106482 (SA2292). Furthermore, fruitful discussions with Gnanavel Vaidyanathan and Markus Olbrich are gratefully acknowledged.

\section{Data availability}
The raw/processed data required to reproduce these findings cannot be shared at this time, as the data also forms part of an ongoing study.

\section{Declaration of generative AI use}
The authors declare that \textbf{NO} generative AI tools were used to prepare the text or images in this work.

\printcredits


\begin{thebibliography}{10}

\bibitem{sugioka2017progress}
Sugioka K.
\newblock Progress in ultrafast laser processing and future prospects.
\newblock {\em Nanophotonics}, 6(2):393--413, 2017.
\newblock \href {https://doi.org/10.1515/nanoph-2016-0004}
  {\path{doi:10.1515/nanoph-2016-0004}}.

\bibitem{shugaev2016fundamentals}
Shugaev MV, Wu~C, Armbruster O, Naghilou A, Brouwer N, Ivanov DS, Derrien TJ-Y,
  Bulgakova NM, Kautek W, Rethfeld B, et~al.
\newblock Fundamentals of ultrafast laser--material interaction.
\newblock {\em MRS Bulletin}, 41(12):960--968, 2016.
\newblock \href {https://doi.org/10.1557/mrs.2016.274}
  {\path{doi:10.1557/mrs.2016.274}}.

\bibitem{lin2025advancing}
Lin Z, Ji~L, and Hong M.
\newblock Advancing manufacturing limits: Ultrafast laser nanofabrication
  techniques.
\newblock {\em Engineering}, 2025.
\newblock \href {https://doi.org/10.1016/j.eng.2025.03.017}
  {\path{doi:10.1016/j.eng.2025.03.017}}.

\bibitem{guo2019ultrafast}
Guo B, Sun J, Lu~Y, and Jiang L.
\newblock Ultrafast dynamics observation during femtosecond laser-material
  interaction.
\newblock {\em International Journal of Extreme Manufacturing}, 1(3):032004,
  2019.
\newblock \href {https://doi.org/10.1088/2631-7990/ab3a24}
  {\path{doi:10.1088/2631-7990/ab3a24}}.

\bibitem{wang2022ultrafast}
Wang D, Wei S, Yuan X, Liu Z, Weng Y, Zhou Y, Xiao T-H, Goda K, Liu S, and Lei
  C.
\newblock Ultrafast imaging for uncovering laser--material interaction
  dynamics.
\newblock {\em International Journal of Mechanical System Dynamics},
  2(1):65--81, 2022.
\newblock \href {https://doi.org/10.1002/msd2.12024}
  {\path{doi:10.1002/msd2.12024}}.

\bibitem{martin2019ultrafast}
Martin AA, Calta NP, Hammons JA, Khairallah SA, Nielsen MH, Shuttlesworth RM,
  Sinclair N, Matthews MJ, Jeffries JR, Willey TM, et~al.
\newblock {Ultrafast dynamics of laser-metal interactions in additive
  manufacturing alloys captured by in situ X-ray imaging}.
\newblock {\em Materials Today Advances}, 1:100002, 2019.
\newblock \href {https://doi.org/10.1016/j.mtadv.2019.01.001}
  {\path{doi:10.1016/j.mtadv.2019.01.001}}.

\bibitem{andreev2017ultrafast}
Andreev SV, Aseev SA, Bagratashvili VN, Vorob'ev NS, Ishchenko AA, Kompanets
  VO, Malinovsky AL, Mironov BN, Timofeev AA, Chekalin SV, et~al.
\newblock Ultrafast transmission electron microscope for studying the dynamics
  of the processes induced by femtosecond laser beams.
\newblock {\em Quantum Electronics}, 47(2):116, 2017.
\newblock \href {https://doi.org/10.1070/QEL16276}
  {\path{doi:10.1070/QEL16276}}.

\bibitem{miloshevsky2022ultrafast}
Miloshevsky G.
\newblock Ultrafast laser matter interactions: modeling approaches, challenges,
  and prospects.
\newblock {\em Modelling and Simulation in Materials Science and Engineering},
  30(8):083001, 2022.
\newblock \href {https://doi.org/10.1088/1361-651X/ac8abc}
  {\path{doi:10.1088/1361-651X/ac8abc}}.

\bibitem{zhigilei2009atomistic}
Zhigilei LV, Lin Z, and Ivanov DS.
\newblock Atomistic modeling of short pulse laser ablation of metals:
  connections between melting, spallation, and phase explosion.
\newblock {\em The Journal of Physical Chemistry C}, 113(27):11892--11906,
  2009.
\newblock \href {https://doi.org/10.1021/jp902294m}
  {\path{doi:10.1021/jp902294m}}.

\bibitem{liu2024review}
Liu H, Xie W, Ding Y, Chen K, Wang S, Huo H, and Yang L.
\newblock Review of molecular dynamics simulations in laser-based
  micro/nano-fabrication.
\newblock {\em Nanoscale}, 16(46):21189--21215, 2024.
\newblock \href {https://doi.org/10.1039/D4NR03305A}
  {\path{doi:10.1039/D4NR03305A}}.

\bibitem{song2023critical}
Song S, Lu~Q, Zhang P, Yan H, Shi H, Yu~Z, Sun T, Luo Z, and Tian Y.
\newblock {A critical review on the simulation of ultra-short pulse laser-metal
  interactions based on a two-temperature model (TTM)}.
\newblock {\em Optics \& Laser Technology}, 159:109001, 2023.
\newblock \href {https://doi.org/10.1016/j.optlastec.2022.109001}
  {\path{doi:10.1016/j.optlastec.2022.109001}}.

\bibitem{yin2021molecular}
Yin CP, Zhang ST, Dong YM, Ye~QW, and Li~Q.
\newblock Molecular-dynamics study of multi-pulsed ultrafast laser interaction
  with copper.
\newblock {\em Advances in Production Engineering \& Management},
  16(4):457--472, 2021.
\newblock \href {https://doi.org/10.14743/apem2021.4.413}
  {\path{doi:10.14743/apem2021.4.413}}.

\bibitem{ivanov2023atomistic}
Ivanov DS, Terekhin PN, Kudryashov SI, Klimentov SM, Kabashin AV, Garcia ME,
  Rethfeld B, and Zavestovskaya IN.
\newblock The atomistic perspective of nanoscale laser ablation.
\newblock In {\em Ultrafast Laser Nanostructuring: The Pursuit of Extreme
  Scales}, pages 65--137. Springer, 2023.
\newblock \href {https://doi.org/10.1007/978-3-031-14752-4_2}
  {\path{doi:10.1007/978-3-031-14752-4_2}}.

\bibitem{valavanis2024laser}
Valavanis AS, Chen C, Grigoropoulos CP, Eliceiri M, Li~J, and Zhigilei LV.
\newblock Laser fluence and film thickness dependence of the mechanisms of
  femtosecond laser ablation of ag films from atomistic simulations and optical
  imaging.
\newblock In {\em High-Power Laser Ablation VIII}, page PC129391E. SPIE, 2024.
\newblock \href {https://doi.org/10.1117/12.3012635}
  {\path{doi:10.1117/12.3012635}}.

\bibitem{xie2025molecular}
Xie L, Li~Y, Wang F, Yao C, and Yuan X.
\newblock Molecular dynamics simulation study on the ablative copper forming
  process at the atomic level by femtosecond laser.
\newblock {\em Journal of Manufacturing Processes}, 150:38--47, 2025.
\newblock \href {https://doi.org/10.1016/j.jmapro.2025.06.043}
  {\path{doi:10.1016/j.jmapro.2025.06.043}}.

\bibitem{amoruso2014ultrashort}
Amoruso S, Nedyalkov NN, Wang X, Ausanio G, Bruzzese R, and Atanasov PA.
\newblock Ultrashort-pulse laser ablation of gold thin film targets: Theory and
  experiment.
\newblock {\em Thin Solid Films}, 550:190--198, 2014.
\newblock \href {https://doi.org/10.1016/j.tsf.2013.10.165}
  {\path{doi:10.1016/j.tsf.2013.10.165}}.

\bibitem{lian2024atomistic}
Lian Y, Jiang L, Sun J, Lin G, and Liang M.
\newblock Atomistic insight on temperature-dependent laser induced ultrafast
  thermomechanical response in aluminum film.
\newblock {\em International Journal of Heat and Mass Transfer}, 231:125809,
  2024.
\newblock \href {https://doi.org/10.1016/j.ijheatmasstransfer.2024.125809}
  {\path{doi:10.1016/j.ijheatmasstransfer.2024.125809}}.

\bibitem{rouleau2014nanoparticle}
Rouleau CM, Shih C-Y, Wu~C, Zhigilei LV, Puretzky AA, and Geohegan DB.
\newblock Nanoparticle generation and transport resulting from femtosecond
  laser ablation of ultrathin metal films: Time-resolved measurements and
  molecular dynamics simulations.
\newblock {\em Applied Physics Letters}, 104(19), 2014.
\newblock \href {https://doi.org/10.1063/1.4876601}
  {\path{doi:10.1063/1.4876601}}.

\bibitem{zhang2021mechanisms}
Zhang Z, Yang Z, Wang C, Zhang Q, Zheng S, and Xu~W.
\newblock {Mechanisms of femtosecond laser ablation of Ni3Al: Molecular
  dynamics study}.
\newblock {\em Optics \& Laser Technology}, 133:106505, 2021.
\newblock \href {https://doi.org/10.1016/j.optlastec.2020.106505}
  {\path{doi:10.1016/j.optlastec.2020.106505}}.

\bibitem{hayder2024effective}
Hayder MM, Moumita TM, Chowdhury S, and Rahman KA.
\newblock Effective electronic properties and coupling for two-temperature
  model-molecular dynamics simulation of ultrafast laser ablation of nickel.
\newblock {\em Molecular Simulation}, 50(14):1140--1151, 2024.
\newblock \href {https://doi.org/10.1080/08927022.2024.2385499}
  {\path{doi:10.1080/08927022.2024.2385499}}.

\bibitem{ganesan2025capturing}
Ganesan H and Sandfeld S.
\newblock Capturing thin-film microstructure contributions during ultrafast
  laser-metal interactions using atomistic simulations.
\newblock {\em Materials \& Design}, page 114224, 2025.
\newblock \href {https://doi.org/10.1016/j.matdes.2025.114224}
  {\path{doi:10.1016/j.matdes.2025.114224}}.

\bibitem{olbrich2016investigation}
Olbrich M, Punzel E, Lickschat P, Wei{\ss}mantel S, and Horn A.
\newblock Investigation on the ablation of thin metal films with femtosecond to
  picosecond-pulsed laser radiation.
\newblock {\em Physics Procedia}, 83:93--103, 2016.
\newblock \href {https://doi.org/10.1016/j.phpro.2016.08.017}
  {\path{doi:10.1016/j.phpro.2016.08.017}}.

\bibitem{ganesan2018parallelization}
Ganesan H, Teijeiro C, and Sutmann G.
\newblock Parallelization comparison and optimization of a scale-bridging
  framework to model cottrell atmospheres.
\newblock {\em Computational materials science}, 155:439--449, 2018.
\newblock \href {https://doi.org/10.1016/j.commatsci.2018.08.055}
  {\path{doi:10.1016/j.commatsci.2018.08.055}}.

\bibitem{ganesan2021quantifying}
Ganesan H, Scheider I, and Cyron CJ.
\newblock Quantifying the high-temperature separation behavior of lamellar
  interfaces in $\gamma$-titanium aluminide under tensile loading by molecular
  dynamics.
\newblock {\em Frontiers in materials}, 7:602567, 2021.
\newblock \href {https://doi.org/10.3389/fmats.2020.602567}
  {\path{doi:10.3389/fmats.2020.602567}}.

\bibitem{ganesan2025modeling}
Ganesan H and Sutmann G.
\newblock Modeling segregated solutes in plastically deformed alloys using
  coupled molecular dynamics-monte carlo simulations.
\newblock {\em Journal of Materials Science \& Technology}, 213:98--108, 2025.
\newblock \href {https://doi.org/10.1016/j.jmst.2024.06.030}
  {\path{doi:10.1016/j.jmst.2024.06.030}}.

\bibitem{norman2012atomistic}
Norman GE, Starikov SV, and Stegailov VV.
\newblock Atomistic simulation of laser ablation of gold: effect of pressure
  relaxation.
\newblock {\em Journal of Experimental and Theoretical Physics}, 114:792--800,
  2012.
\newblock \href {https://doi.org/10.1134/S1063776112040115}
  {\path{doi:10.1134/S1063776112040115}}.

\bibitem{yao2022exploring}
Yao J, Qi~D, Liang H, He~Y, Yao Y, Jia T, Yang Y, Sun Z, and Zhang S.
\newblock Exploring femtosecond laser ablation by snapshot ultrafast imaging
  and molecular dynamics simulation.
\newblock {\em Ultrafast Science}, 2022.
\newblock \href {https://doi.org/10.34133/2022/9754131}
  {\path{doi:10.34133/2022/9754131}}.

\bibitem{thompson2022lammps}
Thompson AP, Aktulga HM, Berger R, Bolintineanu DS, Brown WM, Crozier PS,
  In't~Veld PJ, A~Kohlmeyer, SG~Moore, TD~Nguyen, et~al.
\newblock {LAMMPS-a flexible simulation tool for particle-based materials
  modeling at the atomic, meso, and continuum scales}.
\newblock {\em Computer Physics Communications}, 271:108171, 2022.
\newblock \href {https://doi.org/10.1016/j.cpc.2021.108171}
  {\path{doi:10.1016/j.cpc.2021.108171}}.

\bibitem{rutherford2007effect}
Rutherford AM and Duffy DM.
\newblock The effect of electron--ion interactions on radiation damage
  simulations.
\newblock {\em Journal of Physics: Condensed Matter}, 19(49):496201, 2007.
\newblock \href {https://doi.org/10.1088/0953-8984/19/49/496201}
  {\path{doi:10.1088/0953-8984/19/49/496201}}.

\bibitem{daw1984embedded}
Daw MS.
\newblock Embedded-atom method: Derivation and application to impurities,
  surfaces, and other defects in metals.
\newblock {\em Physical Review B}, 29(12):6443, 1984.
\newblock \href {https://doi.org/10.1103/PhysRevB.29.6443}
  {\path{doi:10.1103/PhysRevB.29.6443}}.

\bibitem{ganesan2021understanding}
Ganesan H, Scheider I, and Cyron CJ.
\newblock {Understanding creep in TiAl alloys on the nanosecond scale by
  molecular dynamics simulations}.
\newblock {\em Materials \& Design}, 212:110282, 2021.
\newblock \href {https://doi.org/10.1016/j.matdes.2021.110282}
  {\path{doi:10.1016/j.matdes.2021.110282}}.

\bibitem{chandran2024studying}
Chandran A, Ganesan H, and Cyron CJ.
\newblock {Studying the effects of Nb on high-temperature deformation in TiAl
  alloys using atomistic simulations}.
\newblock {\em Materials \& Design}, 237:112596, 2024.
\newblock \href {https://doi.org/10.1016/j.matdes.2023.112596}
  {\path{doi:10.1016/j.matdes.2023.112596}}.

\bibitem{grovenor1984development}
Grovenor CRM, Hentzell HTG, and Smith DA.
\newblock {The development of grain structure during growth of metallic films}.
\newblock {\em Acta Metallurgica}, 32(5):773--781, 1984.
\newblock \href {https://doi.org/10.1016/0001-6160(84)90150-0}
  {\path{doi:10.1016/0001-6160(84)90150-0}}.

\bibitem{olbrich2020hydrodynamic}
Olbrich M, Pflug T, W{\"u}stefeld C, Motylenko M, Sandfeld S, Rafaja D, and
  Horn A.
\newblock Hydrodynamic modeling and time-resolved imaging reflectometry of the
  ultrafast laser-induced ablation of a thin gold film.
\newblock {\em Optics and Lasers in Engineering}, 129:106067, 2020.

\bibitem{hirel2015atomsk}
Hirel P.
\newblock {Atomsk: A tool for manipulating and converting atomic data files}.
\newblock {\em Computer Physics Communications}, 197:212--219, 2015.
\newblock \href {https://doi.org/10.1016/j.cpc.2015.07.012}
  {\path{doi:10.1016/j.cpc.2015.07.012}}.

\bibitem{ivanov2003combined}
Ivanov DS and Zhigilei LV.
\newblock {Combined atomistic-continuum modeling of short-pulse laser melting
  and disintegration of metal films}.
\newblock {\em Physical review B}, 68(6):064114, 2003.
\newblock \href {https://doi.org/10.1103/PhysRevB.68.064114}
  {\path{doi:10.1103/PhysRevB.68.064114}}.

\bibitem{iabbaden2022molecular}
Iabbaden D, Amodeo J, Fusco C, Garrelieand F, and Colombier J-P.
\newblock Molecular dynamics simulation of structural evolution in crystalline
  and amorphous cuzr alloys upon ultrafast laser irradiation.
\newblock {\em Physical Review Materials}, 6(12):126001, 2022.
\newblock \href {https://doi.org/10.1103/PhysRevMaterials.6.126001}
  {\path{doi:10.1103/PhysRevMaterials.6.126001}}.

\bibitem{stukowski2009visualization}
Stukowski A.
\newblock {Visualization and analysis of atomistic simulation data with
  OVITO--the open Visualization Tool}.
\newblock {\em Modelling and Simulation in Materials Science and Engineering},
  18(1):015012, 2009.
\newblock \href {https://doi.org/10.1088/0965-0393/18/1/015012}
  {\path{doi:10.1088/0965-0393/18/1/015012}}.

\bibitem{huff2022residual}
Huff M.
\newblock Residual stresses in deposited thin-film material layers for
  micro-and nano-systems manufacturing.
\newblock {\em Micromachines}, 13(12):2084, 2022.
\newblock \href {https://doi.org/10.3390/mi13122084}
  {\path{doi:10.3390/mi13122084}}.

\bibitem{chen2011time}
Chen J, Chen W-K, Tang J, and Rentzepis PM.
\newblock Time-resolved structural dynamics of thin metal films heated with
  femtosecond optical pulses.
\newblock {\em Proceedings of the National Academy of Sciences},
  108(47):18887--18892, 2011.
\newblock \href {https://doi.org/10.1073/pnas.1115237108}
  {\path{doi:10.1073/pnas.1115237108}}.

\bibitem{wu2022ultrafast}
Wu~J, Tang M, Zhao L, Zhu P, Jiang T, Zou X, Hong L, Luo S-N, Xiang D, and
  Zhang J.
\newblock Ultrafast atomic view of laser-induced melting and breathing motion
  of metallic liquid clusters with mev ultrafast electron diffraction.
\newblock {\em Proceedings of the National Academy of Sciences},
  119(4):e2111949119, 2022.
\newblock \href {https://doi.org/10.1073/pnas.2111949119}
  {\path{doi:10.1073/pnas.2111949119}}.

\bibitem{liubchenko2025laser}
Liubchenko O, Antonowicz J, Sokolowski-Tinten K, Zalden P, Minikayev R, Milov
  I, Albert TJ, Bressler C, Chojnacki M, D{\l}u{\.z}ewski P, et~al.
\newblock Laser-induced ultrafast structural transformations in thin fe layer
  revealed by time-resolved x-ray diffraction.
\newblock {\em arXiv preprint arXiv:2506.18730}, 2025.
\newblock \href {https://doi.org/10.48550/arXiv.2506.18730}
  {\path{doi:10.48550/arXiv.2506.18730}}.

\bibitem{zhou2022experimental}
Zhou S and Shen H.
\newblock Experimental and numerical studies on micro-bumps without melting of
  gold films with different thicknesses induced by ultrafast laser.
\newblock {\em Optics Communications}, 514:128178, 2022.
\newblock \href {https://doi.org/10.1016/j.optcom.2022.128178}
  {\path{doi:10.1016/j.optcom.2022.128178}}.

\end{thebibliography}
\end{document}